# A Novel Approach for Object Based Audio Broadcasting

Object Based Audio (OBA) provides a new kind of audio experience, delivered to the audience to personalize and customize their experience of listening and to give them choice of what and how to hear their audio content. OBA can be applied to different platforms such as broadcasting, streaming and cinema sound. This paper presents a novel approach for creating object-based audio on the production side. The approach here presents Sample-by-Sample Object Based Audio (SSOBA) embedding. SSOBA places audio object samples in such a way that allows audiences to easily individualize their chosen audio sources according to their interests and needs. SSOBA is an extra service and not an alternative, so it is also compliant with legacy audio players.

The biggest advantage of SSOBA is that it does not require any special additional hardware in the broadcasting chain and it is therefore easy to implement and equip legacy players and decoders with enhanced ability. Input audio objects, number of output channels and sampling rates are three important factors affecting SSOBA performance and specifying it to be lossless or lossy. SSOBA adopts interpolation at the decoder side to compensate for eliminated samples. Both subjective and objective experiments are carried out to evaluate the output results at each step. MUSHRA subjective experiments conducted after the encoding step shows good-quality performance of SSOBA with up to five objects. SNR measurements and objective experiments, performed after decoding and interpolation, show significant successful recovery and separation of audio objects. Experimental results show that a minimum sampling rate of 96 kHz is indicated to encode up to five objects in a Stereo-mode channel to acquire good subjective and objective results simultaneously.

## I. INTRODUCTION

Mutual interaction of content and audiences has become a key feature of recent programme developments. Audio, as an important part of multimedia contents, plays an important role in creating an interactive medium between content creators and audiences. Therefore, separation of audio sources is an important aspect of digital audio signal processing. A feature of human hearing is that humans can focus their attention on any one sound source out of a mixture of many in real-time (the 'cocktail-party effect'). Multimedia content has been far from having this ability. However, the ability of content selection out of a mixture of different received multimedia sources is a next step forward towards socialising media. Discrimination of audio sources has multiple applications for audiences, so broadcasters aim to equip themselves with such technology to offer new choices to their users. Teleconferencing, remixing applications, on-line gaming, along with filmmaking, music production and the coverage of sport events, are common applications of audio separation. In addition, hearing-impaired viewers can improve their ability to hear more clearly. Object Based Audio (OBA) is a new kind of audio representation, creating flexible audio content so that the front-end users could have more choices while listening; having the flexibility to choose what and how they hear. Actually, OBA is a part of Next Generation Audio (NGA), which seeks to increase accessibility, personalisation and immersion in audio content production.

This paper is organized as follows: Section II considers a review of previous research on OBA broadcasting. Section III outlines the proposed algorithm. The experimental results are presented in Section IV. Finally, discussion and future work are contained in Section V, and Section VI concludes the paper.

## II. A REVIEW OF PREVIOUS RESEARCH ON OBA BROADCASTING

Different approaches have been adopted to discern audio sources. The first approach uses postproduction pure signal processing methods known as Blind Source Separation (BSS). The implementation of this approach is not yet reliable, and the computation cost of typically complex algorithms makes it rare for use in broadcast technology. Other approaches to classify sound events can be used, such as Genetic Algorithm and deep learning methods. In the OBA approach, as described in this paper, audio sources are considered as distinct objects, and pre-production methods are applied to the signal in order to simplify the process of separation at the playback side. Using special, carefully placed microphones in pre-production is also an important factor in separating audio objects. Using meta-data to describe the objects and therefore discriminate them is an approach of growing importance in broadcasting. And, using a range of sensors to cover a football broadcast to allow users navigate within the scene is also a feature of OBA.

Integration of channel-based, object based and scene-based audio has been presented by the MPEG group as new standard called MPEG-H which enables flexible reproduction of 3D sound. Dialogue Enhancement is one of the most promising applications of user interactivity enabled by OBA, supported by MPEG-H. MPEG-H could also be extended to multichannel sound broadcasting services, as in Japan, which has started to use 22.2 channel with the 8K satellite broadcasting. Dolby has also presented its Next Generation Audio, known as AC-4, which supports OBA to enhance the audio experience, including immersive audio and advanced personalisation of the user experience. The Advanced Television Systems Committee (ATSC) has issued ATSC3.0, which includes MPEG-H as a new TV audio system for broadcasting. An interesting project on OBA broadcast is the ORPHEUS audio project, which involves ten European major players, broadcasters, manufacturers and research institutions. It is a pilot end-to-end media chain for audio broadcast. It contains immersive sound, foreground/background control, language selection, and in-depth programme metadata. In addition to audio content, image and video can also be produced in an object-based form. Combination of object-based audio and video could represent a new kind of media content production. In addition to broadcasting, OBA can be utilised in streaming to improve accessibility. Representation of audio sources as distinct objects would be helpful in achieving more efficient coding. Spatial Audio Object Coding (SAOC) is one the most popular standardisations of the MPEG, to encode multiple audio objects at low bitrates.

## III. THE PROPOSED ALGORITHM

In this section, we propose a new algorithm called Sample-by-Sample Object Based Audio (SSOBA) which provides efficient object encoding and decoding with low computation cost; making it applicable to use in broadcasting and streaming. The SSOBA algorithm gets the input from distinct objects and puts them together sample-by-sample in a manner that allows the user to choose among them without corrupting the intelligibility of the audio. One advantage of the proposed algorithm is its efficiency, which maintains the quality of original objects, along with simplicity, that demands lower computation cost and is therefore easy to implement. On the other hand, utilising SSOBA does not change the intelligibility of the encoded audio and audiences do not detect the difference of the new encoded signal, which is an important advantage in broadcasting industry.

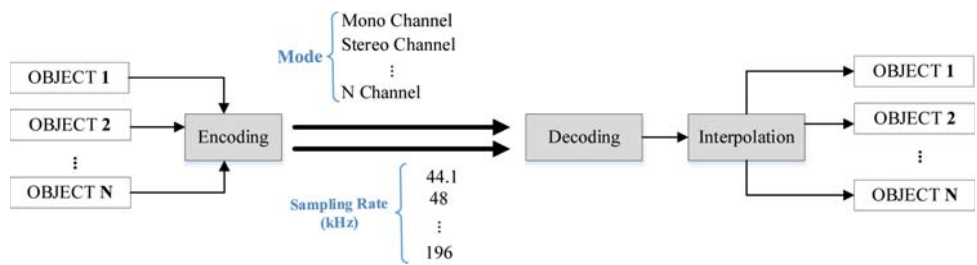

*Figure 1: The general framework of the proposed algorithm **(here N is up to 10)**.*

As an important feature, audio encoded with SSOBA can also be played in legacy players. In fact, SSOBA is an extra service, not an alternative to the current audio players. Therefore, unless users are equipped with players capable of decoding SSOBA, they only miss the opportunity of selecting and separating objects but do not miss the whole signal. This is an important issue in broadcasting, which requires that the audio-visual equipment owned by people reflects the parameters of the macroeconomics of the country, so that broadcasters do not have to force the residents of a region to frequently update their technology.

One way to decrease the slope of economic pressure caused by new technologies is to increase the potential of current technology to its maximum capability. In other words, to develop innovative approaches which can be applied to current systems. SSOBA is such an algorithm. While it presents new services, it is not an alternative to current systems and does not force users to benefit from its properties. Thus audio enhanced with SSOBA can be played on legacy players with no degradation in quality. The general algorithm framework is shown in **Figure 1**. It is noted here that the concept of audio objects used here is the same as audio sources but, to emphasise that they are discrete sources, we use the term object instead.

### A. Encoding & Decoding Process

The algorithm processes sample-by-sample. In fact, same samples of different audio objects (sources) have equal importance and thus implement equal changes. Same samples from all audio sources are encoded and decoded simultaneously. The amount of change of similar samples depends on the their location(index). Three factors limit the performance of the algorithm:

1- Number of input objects to encode
2- Sampling rate (SR) of the input audio objects
3- Number of output audio channels.

Each of the predefined parameters affect the output quality and therefore determine the considerations that should be taken into account in order to reach the required quality of output. If the number of output channels is equal to the number of input audio objects (or sources), SSOBA acts as a lossless algorithm. To the contrary, if the number of input sources is greater than output channels, SSOBA becomes lossy, and in order to compensate for that, interpolation is used. While the algorithm is not limited to any of the above parameters, an optimum point for these three parameters needs to be found that best fits the application. As mentioned, the proposed method is independent of the output channel number layouts, and we applied the algorithm for a 2-channel (stereo) output audio signal, to allow ordinary consumers to use the simple stereo loudspeaker layouts which can be found in almost any systems.

The key concept of SSOBA lies in the simple but important idea of circular shifting. In fact, usually, at any discrete time instant n, each of the input sources and output channels send and receive simultaneous samples respectively. If at any value of n(time), we were to shift same samples and even, occasionally, to remove some of them, we could then trace the change, recover the samples, and reconstruct them at the receiver. Since we assume two-channel stereo for the output signal, if the number of input audio objects (sources) exceeds two, SSOBA will remove some samples from each source in the process of encoding and interpolate them at the receiver. The encoding algorithm is as follows:

*Encode:*
*Input: Original Signal audio file*
*ObjectNo= number of input objects*
*For i=1: ObjectNo*
    *For j=1:length(Input)*
        *K=Mod(sample, ObjectNo)*
        *Encode=circularshift input(j,:) with K steps*
    *end*
*end*
*Output= Encoded signal.*

The decoding process is the reverse of encoding. **Figure 2** shows the process of encoding in SSOBA. In the figure, *Sample ij* means this is the *ith* sample of the *jth* audio object. As illustrated, if, for example, we assume 5 input objects (sources) i.e. *Object Number=6*, according to SSOBA, the first samples of all original object signals are shifted circularly to the size of *mod(1,Object Number)=1*, the second samples are shifted circularly to the size of *mod(2,Object Number)=2* etc. Finally, the fifth samples of all objects are shifted circularly to the size *mod(5, Object Number)=0* i.e. no shift. Then, we separate the output samples according to the desired number of output channels, as shown in the figure.

### B. INTERPOLATION

According to the algorithm presented in the previous section, in order to encode a signal (embed audio objects) in a lossless manner, the number of objects should be equal to or less than the number of channels. If not (number of objects more than the number of channels), the output is lossy, meaning some of the input samples are eliminated. To reconstruct the encoded audio objects at the decoding side, an interpolation process is applied. Interpolation is used to find any values between discrete points. Polynomial interpolation involves finding a polynomial of order $n$ that passes through the $n+1$ point. The interpolated points are created at the same sample numbers as those removed at the encoding side. The disadvantage of direct fitting an $n$th order polynomial at high $n$ numbers is its oscillatory behaviour outside the required interval. The Spline method is a way to find this $n$th order polynomial while avoiding oscillatory behaviour outside the interval. Spline is a composite function formed by n-low-degree polynomials $p_i(x)$ each fitting $f(x)$ in the interval between $x_{i-1}$ and $x_i$, $(i=1,…, n)$:

$$S(x) = \begin{cases} P_1(x) & x_0 \leq x \leq x_1 \\ \vdots & \vdots \\ P_i(x) & x_{i-1} \leq x \leq x_i \\ \vdots & \vdots \\ P_n(x) & x_{n-1} \leq x \leq x_n \end{cases} \quad (2)$$

- For $S(x)$ to be continuous, two consecutive polynomials $P_i(x)$ and $P_{(i+1)}(x)$ must join at $x_i$:

$$P_i(x_i) = P_{i+1}(x_i) = f(x_i) \quad (3)$$

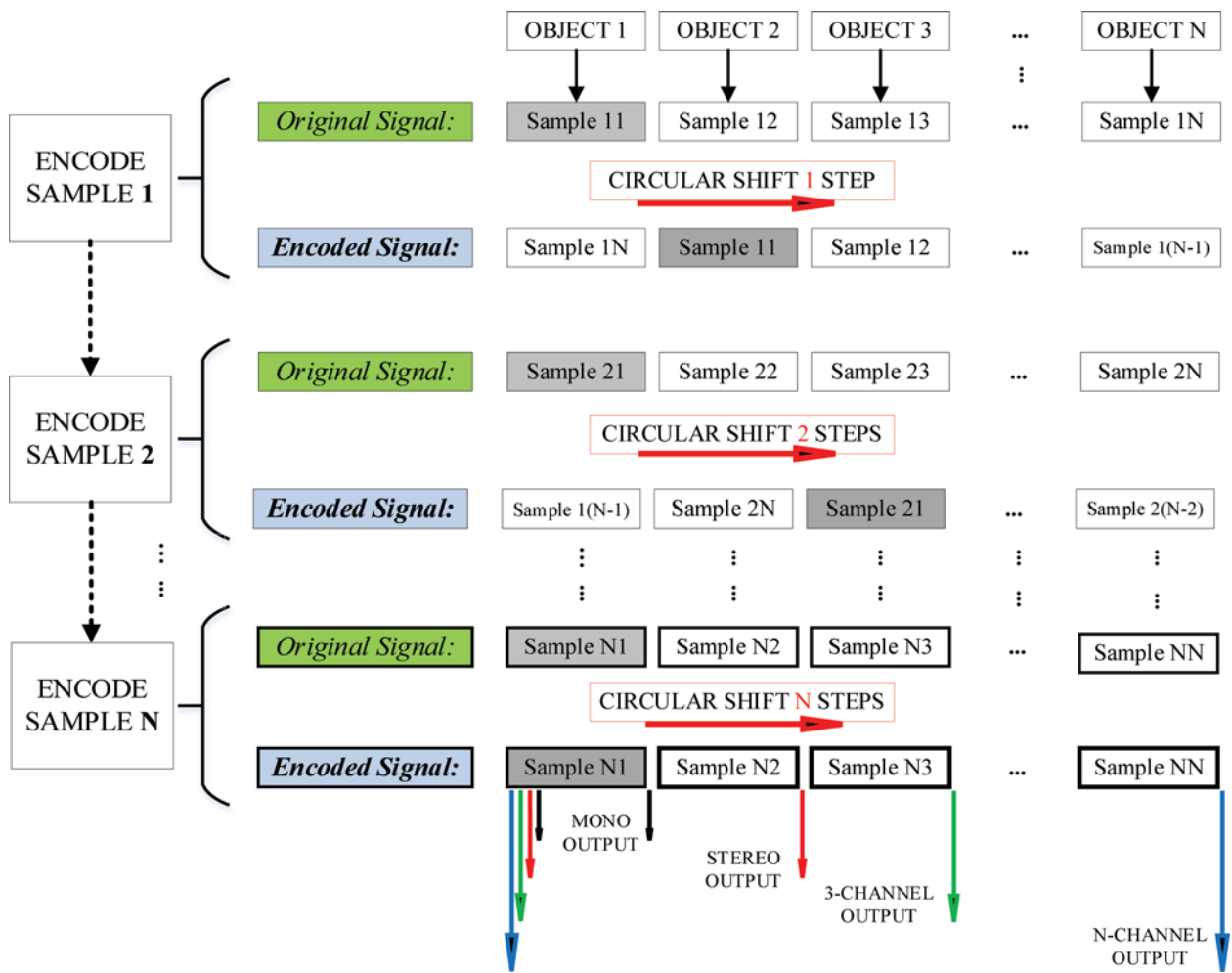

*Figure 2: Sample-by-sample encoding step of SSOBA.*

and also, $P_i(x)$ must pass the two endpoints, i.e.

$$P_i(x_{i-1}) = f(x_{i-1}), \quad P_i(x_i) = f(x_i) \quad (4)$$

- In order to smooth $S(x)$, $P_i(x)$ (i=1,…,n) should have same derivatives at the points they join i.e.,

$$P_i^{(k)}(x_i) = P_{i+1}^{(k)}(x_i). \quad (5)$$

**Linear Spline:** if we consider $P_i(x)=a_i x+b_i$ with two parameters, $a_i$ and $b_i$ are detected through following two equations:

$$P_i(x) = a_i x_i + b_i = f(x_i),$$
$$P_i(x_{i-1}) = a_i x_{i-1} + b_i = f(x_{i-1}) \quad (6)$$

**Quadratic Spline:** if we consider $P_i(x)=a_i x^2+b_i x+c_i$ with three parameters, $a_i$, $b_i$ and $c_i$ are detected through following three equations:

$$P_i(x_i) = f(x_{i-1}), \quad P(x_{i-1}) = f(x_{i-1}),$$
$$P_i^{(1)}(x_i) = P_{i+1}^{(1)}(x_i) \quad (7)$$

**Cubic Spline:** finally, if we consider $P_i(x)=a_i x^3+b_i x^2+c_i x+d_i$ with four parameters, then following four equations are applied [1]:

$$P_i(x_i) = f(x_i), \quad P(x_{i-1}) = f(x_{i-1}),$$
$$P_i^{(1)}(x_i) = P_{i+1}^{(1)}(x_i), \quad P_i^{(2)}(x_i) = P_{i+1}^{(2)}(x_i) \quad (8)$$

Since the cubic spline produces better interpolation and a warmer audio signal, we have selected this option to recover and reconstruct the output objects.

## IV. EXPERIMENTAL RESULTS

We evaluate the performance of encoding, decoding and interpolation through two different assessment categories; *subjective* and *objective* tests.

### A. Subjective tests

Since audio is a phenomenon, which can be truly evaluated through direct hearing, subjective tests are applied after the *encoding* step to assess how much the encoding process has changed the intelligibility of the audio. To evaluate encoding performance, the MUSHRA method is applied. Since the writing of this paper coincided with the outbreak of COVID-19, we performed subjective tests over the Internet, as so-called web-based measurements, using webMUSHRA while maintaining compliance with ITU-R Recommendation BS.1534 (MUSHRA). The illustration of a sample webMUSHRA session test, as exhibited to the assessors, is depicted in **Figure 3**. As shown, each session test includes eleven audio files which could be switched easily and evaluated by comparison to the Reference signal. The Eleven signals contain Reference, Anchor-3.5, Anchor-7, 3-Objects, 4-Objects, 5-Objects, 6-Objects, 7-Objects, 8-Objects, 9-Objects and 10-Objects signals. Anchor-3.5 and Anchor-7 are obtained through low-pass filtering the Reference signal by 3.5 kHz and 7 kHz edges respectively. The loudness of audio test files was configured according to the ITU-R BS.1770-3 Loudness standard. As mentioned earlier, three parameters affect the output quality of the SSOBA: *Input Object Numbers*, *Sampling Rate* and *Output Channel Number*.

For consistency and ease of use, output is considered fixed, as two-channel (stereo), and subjective tests are carried out to evaluate the impact of sampling rate and the number of input objects. Five tests were programmed, each with different sampling rates. Twenty participants took part in the MUSHRA tests. The MUSHRA scores are shown at **Figure 4**. The results show that SSOBA improves with sampling rate. In fact, the higher the sampling rate, the better the scores. As illustrated, the minimum accepted sampling rates for three-objects, four-objects and five-objects are 48kHz, 88.2kHz and 96kHz respectively. Additional sampling rates could have been selected but there are limits due to applications. The MUSHRA scores show that, although SSOBA disrupts the normal sequence of input audio samples, since the samples are made tens of thousands times per second, the disruption of sample sequence is not sensed by the listeners.

### B. Objective test

Objective tests are carried out to evaluate the reconstruction of the output audio objects and are assessed through Signal-to-Noise Ratios (SNR). SNR shows the performance of the encode-decode algorithm and interpolation step. Unless the number of the input objects exceeds the number output audio channels, the SSOBA is a lossless algorithm, otherwise, it is lossy. Again, here we consider two-channel (stereo) as our output. Therefore, in cases where the input objects exceed two, we are using the algorithm in lossy manner. Therefore, some samples are eliminated in the process of encoding, and need to be recovered at the decoding side through interpolation. SNR indicates how similar decoded and cubic spline interpolated audio objects (sources) are to their original versions at the input. SNR is evaluated as follows:

$$SNR = 20\log\frac{std(y)}{std(y-x)} \qquad (9)$$

$$std = \sqrt{\frac{\sum_N (x_i - \mu)^2}{N}} \qquad (10)$$

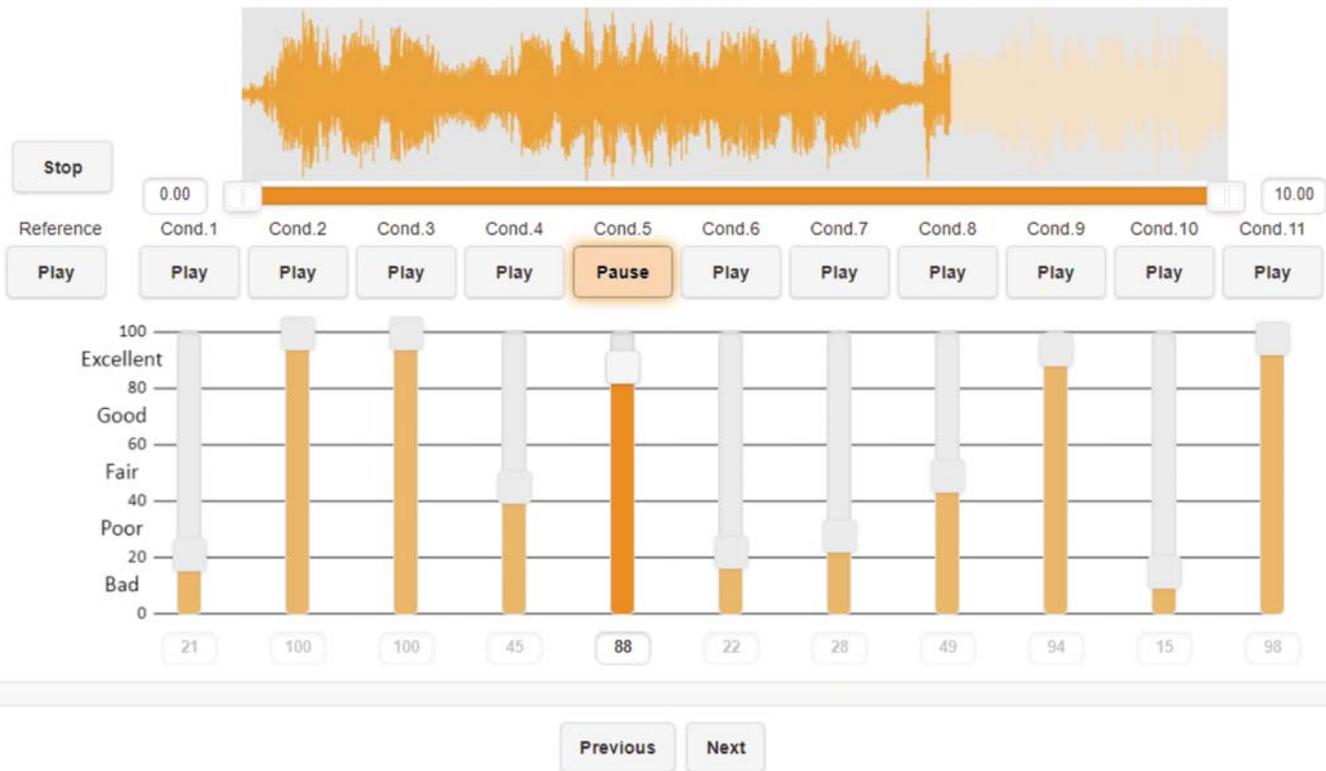

*Figure 3: Screenshot of SSOBA MUSHRA implementation presented to the assessors.*

Where *y* is a decoded and interpolated audio object at the decoder and x is the original audio object (source) before encoding. Of course, we only consider SNR for more than two objects, since below that SSOBA acts without loss and SNR has its ideal value (∞). Since the results repeat, we only consider a class of five input objects and compute the SNR.

*Figure 5* shows the behaviour of each object in a five-object class changing with sampling rate. Again, sampling rate improves SSOBA performance. As illustrated, objects of one class behave similarly. Due to the similar behaviour of audio objects with SR, we now consider SNR of the first object in each case in *Figure 6*. Obviously, the increase of SR of the input audio object improves the distinction of reconstructed output objects.

It is noted that an SNR of 30dB is effectively a clean signal. Listeners will barely notice any impairment where SNR is better than 20dB. Therefore, as SNR shows, in a 3-Object case, the minimum SR is 32 kHz in order to get SNR above 20dB. Above 48 kHz, the output SNRs are good enough. 4-Objects, 5 Objects and 6-Objects need at least a 44.1 kHz sampling rate to recover the Stereo-mode encoded objects with SNR above 20dB. On the other hand, 7-Objects, 8-Objects and 9-Objects need a sampling rate of 64 kHz to achieve SNR better than 20dB. 96 kHz could be a base sampling rate for 10-Object encoding in Stereo-mode to recover objects with more than 20dB SNR at the decoder. Generally speaking, 96 kHz is a recommended sampling rate to encode up to five objects in a Stereo-mode channel to acquire good subjective and objective results simultaneously. It is worth noting that, although the standard audio sampling rate is 48 kHz, the movement of broadcasters towards NGA, including MPEG-H and Dolby Atmos, which covers 22% of continuous UHD, according to UHD Service Tracker B2C, may require high-resolution (HD) audio, which commonly refers to SR greater than 48 kHz. This is the reason why, SRs greater than 48 kHz are considered in this paper.

### V. DISCUSSION AND FUTURE WORK

One advantage of SSOBA is its efficiency, which keeps the quality of original objects, along with simplicity, which involves less computation cost and is therefore easy to implement. On the other hand, adoption of SSOBA does not change the intelligibility of the encoded audio which is an important requirement in the broadcasting industry. An important feature of SSOBA that it does not demand any special extra hardware in the production and distribution chain. It can be easily implemented by application at the output end of audio mixers without any interference with the mixing process. Also, SSOBA does not change the normal MPEG Transport Stream (TS) packetization in the broadcast chain. It is applied before TS encoding, and after TS decoding at the receiver, therefore, no complexity is added to the system.

*Figure 7* shows a broadcast chain in which SSOBA could be applied. SSOBA

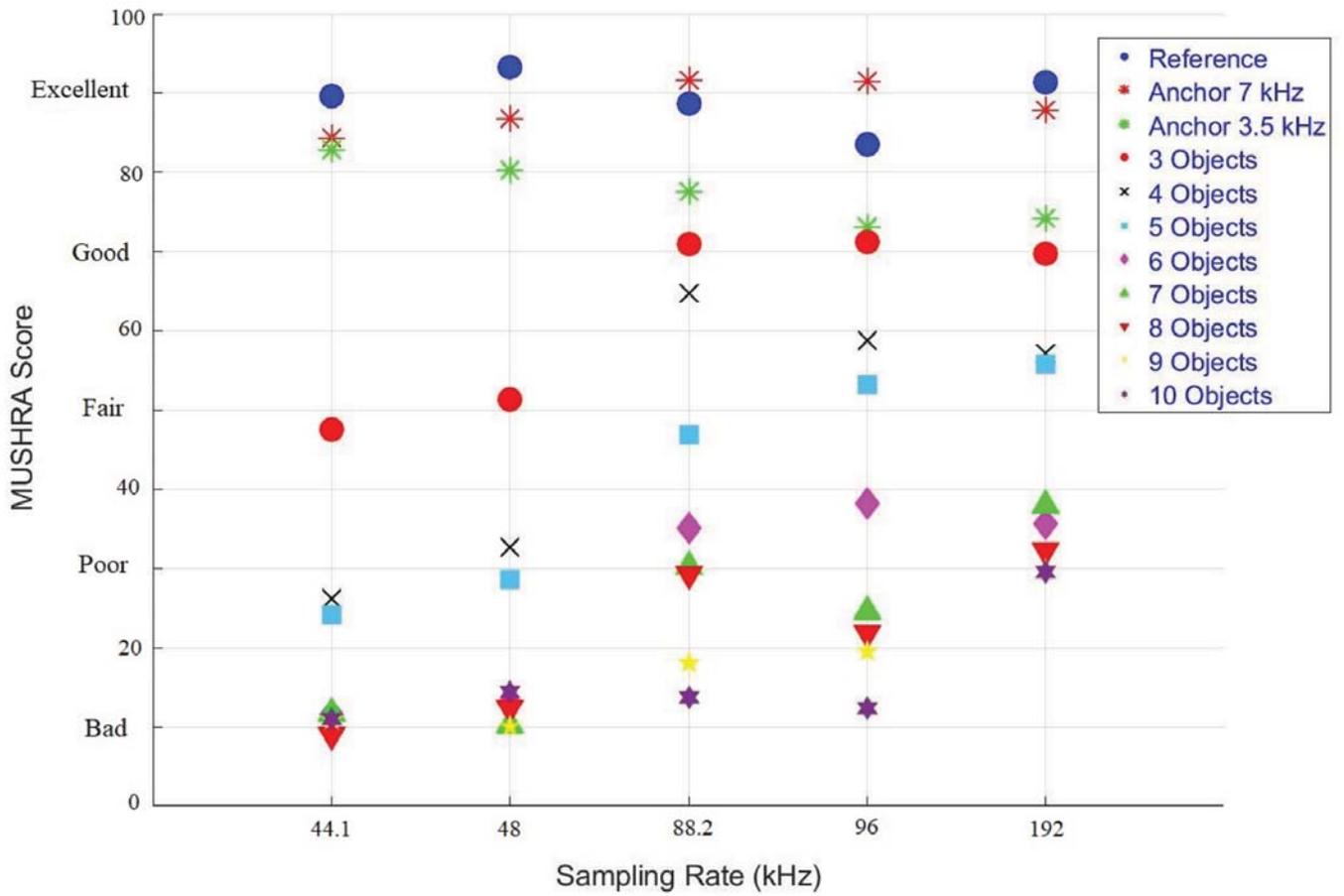

*Figure 4:* The MUSHRA scores of nine audio recordings with 95% confidence interval.
The anchor is obtained by filtering down-mix signal with 3.5 kHz and 7 kHz low-pass filter.

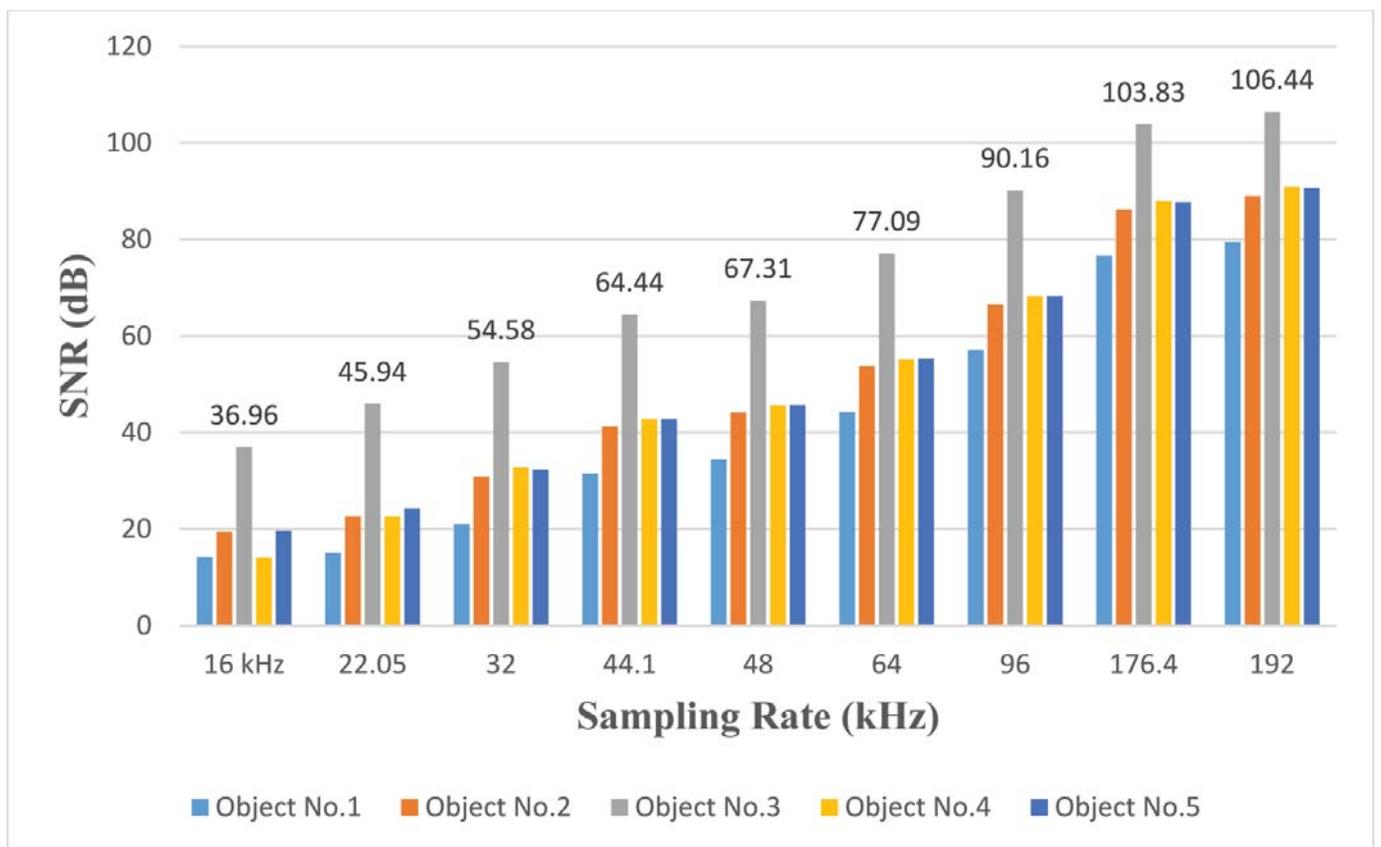

*Figure 5:* SNR results of decoded and interpolated objects of a five-object class in Stereo Encoded Output.

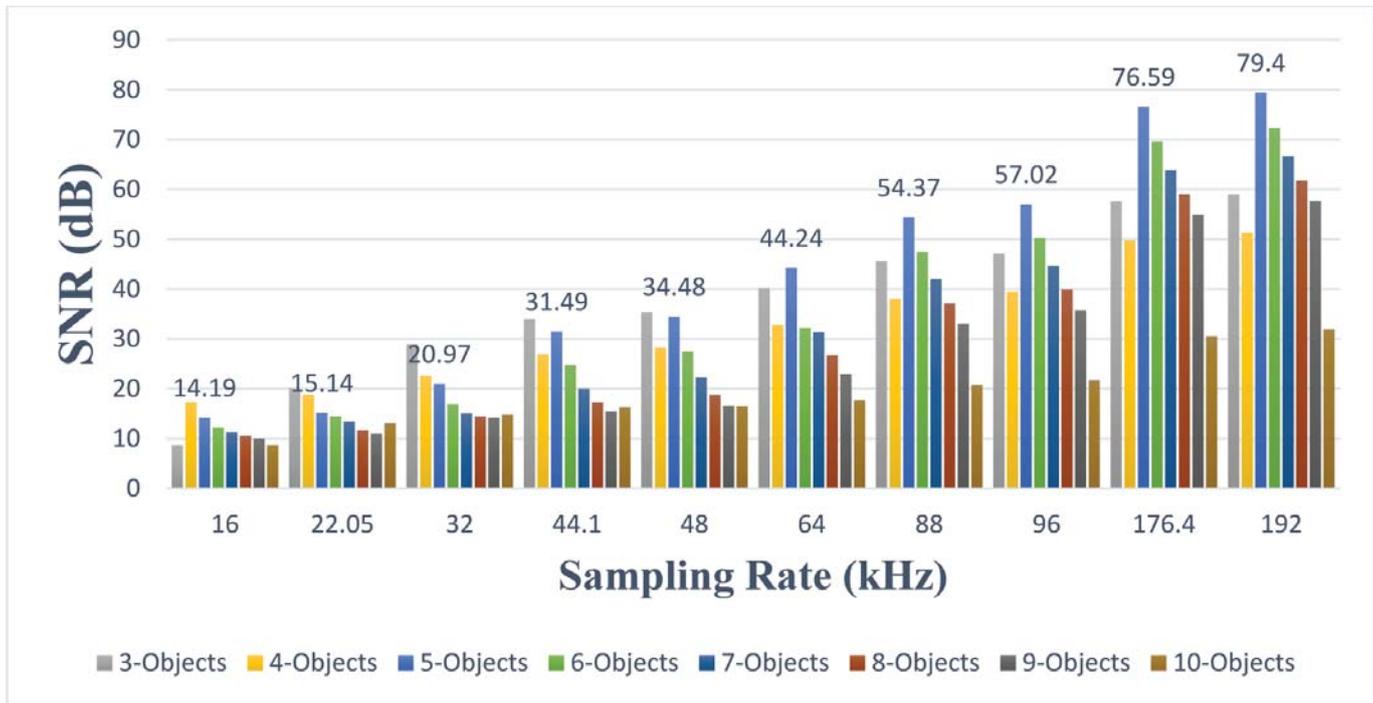

*Figure 6:* SNR results of decoded and interpolated objects in Stereo Encoded Output. The values are considered only for the first object in each class

can be easily implemented in digital smart TVs by the simple addition of App capable of decoding SSOBA, and without the need to change CODEC. In other cases, digital set-top boxes can similarly, be simply updated. In future work, we will concentrate on how to specify other descriptive properties of objects such as spatial placement, loudness, etc. using SSOBA. Generally speaking, 96 kHz is a recommended sampling rate to encode up to five objects in a Stereo-mode channel to simultaneously acquire good subjective and objective results.

## VI. CONCLUSION

In this paper, we proposed SSOBA as an innovative Object Based Audio algorithm. SSOBA processes input audio objects sample-by-sample in a manner which is efficient, simple and computationally economical to apply.

The greatest advantage of SSOBA is that it does not add any special extra hardware to the broadcasting chain, and it is therefore easy to implement and equip legacy players and decoders with its features. An important advantage is that audio encoded with SSOBA can also be played by legacy players. In such cases, the users only miss the opportunity of selecting and separating sources but do not miss the required signal.

The number of input audio objects depends directly on the number of output channels and sampling rate.

In this paper, we focused on two-channel (stereo) output which is currently popular in most audio production. Subjective tests of MUSHRA along with objective tests of SNR showed that an increase in sampling rate significantly improves both subjective and objective results.

Experimental results show that a minimum sampling rate of 96 kHz is a recommended to encode up to five objects in a Stereo-mode channel to acquire good subjective and objective results simultaneously.

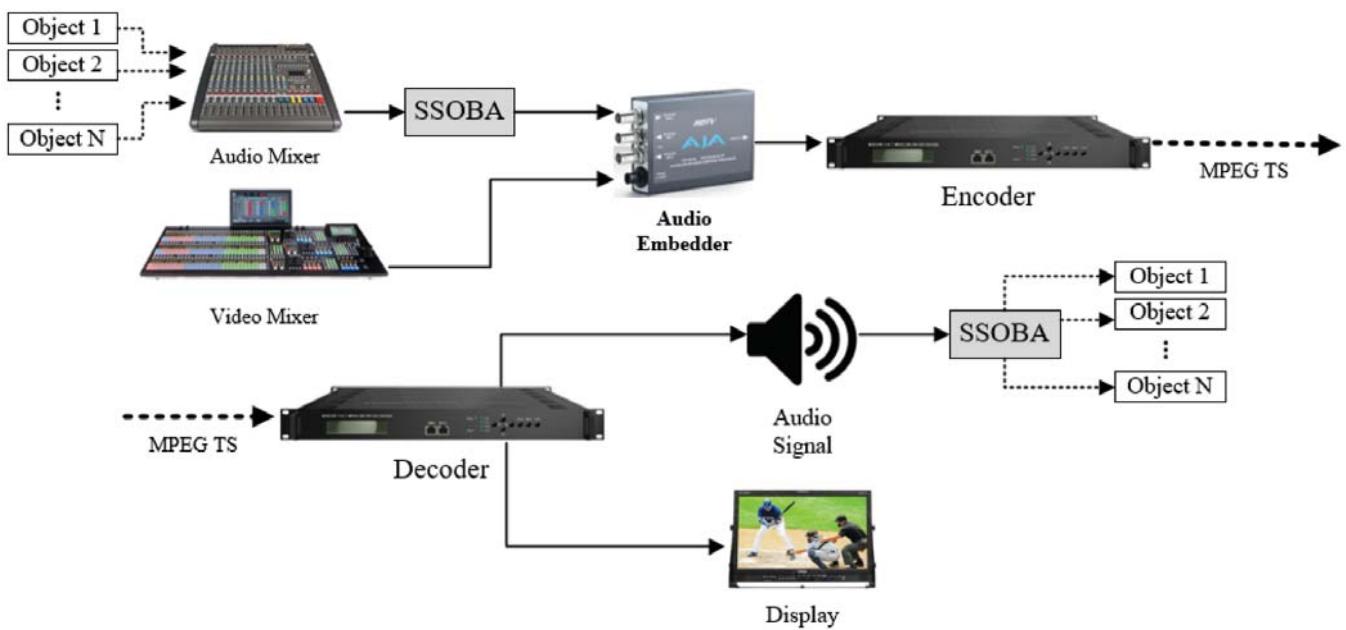

*Figure 7:* Position of SSOBA in Broadcast Chain.

Author:

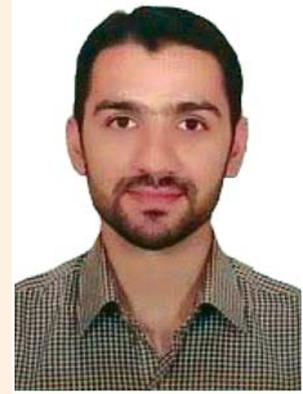

Mohammad Reza Hasanabadi works as a researcher at IRIB R&D. He received His M.S. degree in Audio Engineering from IRIB UNIVERSITY in 2016. He is currently pursuing a Ph.D. in Electrical Engineering at Shahid Beheshti University (SBU), Tehran, Iran. He joined IRIB R&D in 2019, since when he has been focusing on audio related broadcast projects. His research interests include audio and speech processing, speech enhancement and estimation theory. ■